\documentclass[reprint,tightenlines,showpacs,amsmath,amssymb,aps,pra,superscriptaddress]{revtex4-1}

\usepackage{graphicx}
\usepackage{subfigure}
\usepackage{amsmath}
\usepackage{color}

\begin{document}

\title{Optimal control of quantum gates in an exactly solvable non-Markovian open quantum bit system}
\author{Jung-Shen Tai}
\affiliation{Department of Physics and Center for Theoretical
  Sciences, National Taiwan University, Taipei 10617, Taiwan }
\affiliation{Center for Quantum Science and Engineering and National
  Center for Theoretical Physics, National Taiwan University, Taipei
  10617, Taiwan}
\author{Kuan-Ting Lin}
\affiliation{Department of Physics and Center for Theoretical
  Sciences, National Taiwan University, Taipei 10617, Taiwan }
\author{Hsi-Sheng Goan}
\email{goan@phys.ntu.edu.tw}
\affiliation{Department of Physics and Center for Theoretical
  Sciences, National Taiwan University, Taipei 10617, Taiwan }
\affiliation{Center for Quantum Science and Engineering and National
  Center for Theoretical Physics, National Taiwan University, Taipei 10617, Taiwan}
\date{\today}

\begin{abstract}
We apply quantum optimal control theory (QOCT) to an exactly solvable
non-Markovian open quantum bit (qubit) system to achieve state-independent
quantum control and construct high-fidelity quantum gates for moderate
qubit decaying parameters. An important quantity, improvement $\mathcal{I}$,
is proposed and defined to quantify the correction of gate errors due
to the QOCT iteration when the environment effects are
taken into account. With the help of the exact dynamics, we explore
how the gate error is corrected in the open qubit system and determine the conditions for significant improvement. The model adopted in this paper can be implemented experimentally in realistic systems such as the circuit QED system.
  
\end{abstract}


\maketitle

\section{Introduction}
Quantum optimal control theory (QOCT) which
incorporates the optimal control theory with the quantum theory is a
powerful tool and
has attained various physical
achievements
\cite{PhysRevA.37.4950,Kosloff89,TannorBook,PhysRevLett.89.188301,PhysRevLett.99.170501,maximov:184505,PhysRevA.75.012302,PhysRevA.79.060306,PhysRevA.83.053426,Brif10}. It
has also been introduced to quantum gate control to obtain the optimal
control pulses in quantum gate operations. In the literature, quantum
gate control employing QOCT in closed or open systems are studied
\cite{PhysRevLett.89.188301,PhysRevLett.99.170501,PhysRevA.75.012302,PhysRevA.79.060306,JPhysB.40.S75,PhysRevLett.102.090401,PhysRevA.78.012358,*PhysRevB.78.165118,*wenin:084504,*PhysRevB.79.224516,*PhysRevA.74.022319,0295-5075-87-4-40003,PhysRevLett.101.010403,PhysRevLett.104.040401,Hwang2012}. However,
in most investigations where the environment effect is taken into
account, the dynamics are often derived perturbatively, involving Born
\cite{breuer2002theory,PhysRevA.59.1633,Meier99,Breuer01,Yan98,Schroder06,Ferraro09,Shibata77,Kleinekathofer04,Liu07,Sinayskiy09,Mogilevtsev09,Haikka10,Ali10,Chen11,Goan11}
or Born-Markov approximations
\cite{Scully97,carmichael1998statistical,Gardiner00,Barnett02,Milburn08}.
Despite the broad
applicability of the perturbative master equation, the approximations
made in the derivation results in unwanted intrinsic error, which in
turn contributes to the gate error as the pulse sequence for the gate
operation is obtained through the approximated master equation. 
In cases where the models can be exactly solved, resorting to the
exact dynamics can help reduce these possible intrinsic errors.

In this paper, we adopt an exact master equation of a qubit
\cite{breuer2002theory,PhysRevA.54.R3750,PhysRevA.55.2290,PhysRevA.58.1699}
and
combine it with QOCT based on the Krotov iteration method \cite{krotov1996global,TannorBook,PhysRevLett.89.188301,maximov:184505,PhysRevA.66.053619}
to find the optimal control pulse for state-independent
single-qubit gate control in a general non-Markovian environment with
an arbitrary spectral density.
To be specific, the model we consider is a qubit linearly coupled to a
dissipative zero-temperature environment through a qubit lowering operator
$\sigma_-$.  The exact master equation for this model can be
derived from either the pseudo mode method
\cite{PhysRevA.55.2290,breuer2002theory} or the quantum state diffusion equation
\cite{PhysRevA.58.1699,Diosi1997569,*PhysRevLett.82.1801,*PhysRevLett.74.203}.  
Reasonable trends of gate error for this model 
under various conditions are observed and discussed.
Moreover, if the bath spectral density is chosen to be a Lorentzian type,
this dissipative qubit model can be shown to be equivalent
to the damped Jaynes-Cummings model describing the coupling of a qubit 
to a single cavity mode which in turn is coupled to a Markovian
reservoir \cite{PhysRevA.55.2290,breuer2002theory}. Thus our optimal
control results would also have direct 
applications, for example, to
superconducting circuit quantum electrodynamic (QED) systems
\cite{PhysRevA.69.062320,Wallraff2004,Circuit_QED_review,Sarovar05}
that are described very well by the damped 
Jaynes-Cummings model and are controlled relatively easily by external
fields.

Another important property we wish to investigate is whether or not in open
systems, QOCT is able to correct the gate error due to
the environment effect. A quantity, improvement $\mathcal{I}$, is defined to
quantify such correction. For a system where the improvement is large,
including the environment effect becomes essential to the control
problem. Whereas for a system with negligible improvement, the optimal
control pulse developed while the system is considered closed would
suffice. We further explore the region of parameters where  significant
improvement is achieved, and find that improvement is in close
relation to the structure of the environment. 
We note here that in cases where the open quantum system models are not
exactly solvable, one will have to turn to the perturbative master
equation approaches for the optimal control solution \cite{JPhysB.40.S75,PhysRevLett.102.090401,PhysRevA.78.012358,*PhysRevB.78.165118,*wenin:084504,*PhysRevB.79.224516,*PhysRevA.74.022319,0295-5075-87-4-40003,PhysRevLett.101.010403,PhysRevLett.104.040401,Hwang2012}. However, 
in certain models where the exact master
equations are available, our present treatment bears the advantage of ruling out the intrinsic errors due to the perturbative dynamics.


\section{Model and method}
\subsection{Model and exact master equation}

Only a few 
non-Markovian open quantum system models can be exactly
solved
\cite{PhysRevA.55.2290,PhysRevLett.100.180402,PhysRevA.50.3650,PhysRevA.59.1633,PhysRevD.45.2843,Tu08,Goan10,PhysRevLett.105.240403,PhysRevA.58.1699},
and exact dissipative models of a two-level system are even fewer.  
The total Hamiltonian $H_{\text{tot}}$ of the two-level qubit model we consider
consists of three parts
\cite{PhysRevA.58.1699,PhysRevA.55.2290,breuer2002theory} (set $\hbar=1$):
\begin{align}
&H_{\text{qbit}}=\frac{\omega_{0}}{2}\sigma_{z},\notag\\
&H_{\text{bath}}=\sum_{\lambda}\omega_{\lambda}a_{\lambda}^{\dagger}a_{\lambda},\notag\\
&H_{\text{int}}=\sum_{\lambda}(g_{\lambda}^{*}La_{\lambda}^{\dagger}+g_{\lambda}L^{\dagger}a_{\lambda}).
\end{align}
Here $\omega_{0}$ is the qubit transition frequency; $\sigma_{z}$
is the Pauli-$Z$ matrix; and $a_{\lambda}$, $a_{\lambda}^{\dagger}$ are
the creation and annihilation operator for the bath oscillator with
eigenfrequency 
$\omega_{\lambda}$. 
In this exactly solvable model, the qubit is linearly coupled to the
zero-temperature environment 
through the Lindblad operator $L=\sigma_{-}$ with coupling constant
$g_{\lambda}$. 
We choose the qubit transition frequency as the time-dependent control
parameter, $\omega_{0}\rightarrow\omega_{0}+\epsilon(t)\equiv\omega_{0}(t)$.
In real experiments, $\omega_{0}$ is often tunable and is a
possible agent of external control.
The exact master equation reads \cite{PhysRevA.58.1699,PhysRevA.55.2290,breuer2002theory}
\begin{align}
\label{eq: master equation}
&\dot{\rho}(t)=-\frac{i\omega_{0}(t)}{2}\left[\sigma_{z},\rho(t)\right]+2\text{Re}\left[F(t)\right]\notag\\
&\times\left(\sigma_{-}\rho(t)\sigma_{+}-\frac{1}{2}\left\{ \sigma_{+}\sigma_{-},\rho(t)\right\} \right)+i\text{Im}\left[{F(t)}\right]\left[\sigma_{+}\sigma_{-},\rho(t)\right],
\end{align}
where $\text{Re}[\cdots]$ and $\text{Im}[\cdots]$ stand for the real
and imaginary parts of a complex function, 
\begin{equation}
F(t)=\int_0^t c(t,s)f(t,s)ds
\label{eq:F}
\end{equation} 
satisfying the differential equation
\begin{equation}
\partial_tf(t,s)=\{i\left(\omega_0+\epsilon(t)\right)+F(t)\}f(t-s)
\label{eq:F_diff_eq}
\end{equation} 
and the bath correlation function is defined as 
\begin{align}
c(t-s)&\equiv\sum_{\lambda}\left|g_{\lambda}\right|^{2}e^{-i\omega_{\lambda}(t-s)}\notag\\
&\rightarrow\int_{0}^{\infty}d\omega J(\omega)e^{-i\omega(t-s)}, 
\label{eq:bath_CF}
\end{align}
where we have taken the continuum limit and $J(\omega)$ is the
environment spectral density. 
Equation (\ref{eq:F_diff_eq}) is a nonlocal
integro-differential equation and is not easy to solve for a general
bath spectral density and thus to incorporate 
within the framework of QOCT.
One important observation to deal with this time-nonlocal 
equation 
is to express the bath correlation function in a multi-exponential
form \cite{Meier99,Yan01,Kleinekathofer04,Kleinekathofer06,Hwang2012},
\begin{equation}
c(t-s)=\sum_{j}p_{j}e^{q_{j}(t-s)}=\sum_{j}c_{j}(t-s),
\end{equation} 
where $p_j$ and $q_j$ are complex constants and can be found
by numerical methods.
Then we see from Eqs.~(\ref{eq:F}) and (\ref{eq:F_diff_eq})
that the relevant function $F(t)$ in Eq.~(\ref{eq: master equation}) satisfies
$F(t)=\sum_{j}F_{j}(t)$ and 
\begin{align}
\label{eq:partialF}
\partial_{t}F_{j}(t)=&p_{j}+F_{j}(t)\left[q_{j}+i\left[\omega_{0}+\epsilon(t)\right]+\sum_{k\neq j}F_{k}(t)\right]\notag\\
+&F_{j}^{2}(t), 
\end{align} 
along with the initial condition $F_j(0)=0$. Equation
(\ref{eq:partialF}) forms a set of
coupled time-local equations that 
yield a simple, fast and stable iterative scheme to 
incorporate with the Krotov QOCT method that we will employ.  For the
convenience of numerical computation, we treat the density matrix as a
column vector $\rho^{c}$ and  Eq. (\ref{eq: master equation}) can be put in
the form $\dot{\rho}^{c}(t)=\Lambda(t)\rho^{c}(t)$. The propagator
$\mathcal{G}(t)$ is defined such that
$\rho^{c}(t)=\mathcal{G}(t)\rho^{c}(0)$ and can be viewed as a
state-independent gate operation. The differential equation for
$\mathcal{G}(t)$ is $\dot{\mathcal{G}}(t)=\Lambda(t)\mathcal{G}(t)$
and $\mathcal{G}(t)$ is identity when $t=0$. 

\subsection{Krotov's method of optimal control theory}

In QOCT, it is necessary to define a quantity, or the cost function,
we wish to maximize or minimize after each iteration
\cite{krotov1996global,TannorBook,PhysRevLett.89.188301,maximov:184505,PhysRevA.66.053619}.
In open system gate control, this quantity corresponds to the gate error defined at the final gating time $t_f$ \cite{maximov:184505,PhysRevA.78.012358,*PhysRevB.78.165118,*wenin:084504,*PhysRevB.79.224516,*PhysRevA.74.022319},
\begin{equation}
\mathcal{E}\equiv\frac{1}{2\mathcal{N}}\text{Tr}\left\{ [\mathcal{O}-\mathcal{G}(t_{f})]^{\dagger}[\mathcal{O}-\mathcal{G}(t_{f})]\right\},
\end{equation} 
where $\mathcal{O}$ is the control target to be specified and
$\mathcal{N}$ is the dimension of $\mathcal{G}(t)$ in the column vector
representation. 
This error $\mathcal{E}$ or fidelity $(1-\mathcal{E})$ definition can be mapped to the trace fidelity commonly used in closed systems when the dynamics becomes unitary.
For the dissipative two-level model with control over the $\sigma_{z}$
term, we perform $Z$-gate and identity gate control. For the $Z$-gate control, 
the target $\mathcal{O}_{z}$ in the column vector representation 
is defined as $\text{diag}\left(1,-1,-1,1\right)$ and for identity-gate
control $\mathcal{O}_{I}=I_{\mathcal{N}}$, where $I_{\mathcal{N}}$ is the
identity matrix in the column vector representation.  

The update algorithm of the optimization iteration based on the
Krotov method is as follows
\cite{krotov1996global,TannorBook,PhysRevLett.89.188301,maximov:184505,Hwang2012}: (1) An
admissible initial control $\epsilon^{0}(t)$ is constructed either by
guess or experience. Find the trajectory $\mathcal{G}^{(0)}(t)$ by
intergrating the equation of motion along with the initial condition
using the control $\epsilon^{(0)}$. (2) An auxiliary backward
propagator $\chi(t)$ is found by integrating the differential equation
$\dot{\chi}(t)=\Lambda^{\dagger}\chi(t)$ with its boundary condition 
$\chi(t_{f})=[\mathcal{O}-\mathcal{G}(t_{f})]/2\mathcal{N}$.
(3) Solve the equation of motion for $\mathcal{G}(t)$
  and the control update rule 
\begin{equation*}
\epsilon=\epsilon^{(0)}+2\lambda\mbox{Re}\left(\mbox{Tr}\left[\chi^{\dagger}\left.\frac{\partial\Lambda}{\partial\epsilon}\right|_{\epsilon^{(0)}}\mathcal{G}\right]\right)
\end{equation*}
self-consistently to yield the updated control and propagator
$\epsilon^{(1)}$ and $\mathcal{G}^{(1)}$ for a small enough $\lambda$
to ensure the monotonic convergence of the algorithm. (4) Substitute
$\epsilon^{(1)}$ and $\mathcal{G}^{(1)}$ for $\epsilon^{(0)}$ and
$\mathcal{G}^{(0)}$  in step (1) and repeat steps (1) to (3) until the
error converges to a saturated value (a preset error threshold is reached or a given number of iterations has been performed).

We constrain the control parameter---in our case the time-dependent
transition frequency---to an allowable range. In real experiments,
there exists an attainable range of qubit frequency determined by the
external control agent and the physical system. Beyond this range, the
control is unattainable or simply destroys the original system. An
example can be the critical magnetic field in the superconducting
circuit QED system \cite{tinkham2004introduction,PhysRevA.69.062320,Wallraff2004,Circuit_QED_review,Sarovar05}. 
Thus for most of the results presented here, 
we set the range to be $0\leq\omega_{0}(t)\leq2\omega_{0}$, i.e., $\left|\epsilon(t)\right|\leq\omega_{0}$. The value of $\omega_{0}$ is determined by the actual physical system implementing this model. 
We will present results with large range control  
$\left|\epsilon(t)\right|\leq 20\omega_{0}$ in Sec.~\ref{sec:suppression}.

\subsection{Improvement}

In our model, the optimal pulses for $Z$-gates and identity gates in closed systems can be obtained straightforwardly. An important question to be addressed is, how much can QOCT improve the gate fidelity in an open quantum system, given that we take the ideal closed system optimal pulse as our initial guess. For a fixed gating time $t_{f}$ and a constant magnitude control pulse, $\omega_{0}(t)=\omega_{0}+\epsilon(t)=n\pi/t_{f}$ where $n$ is even for the identity gate and odd for the $Z$-gates. We take this ideal closed system pulse as the initial guess for optimal control in open systems. Define the quantity, improvement $\mathcal{I}$: 
\begin{equation}
{
\mathcal{I}\equiv\log_{10}{\left(\frac{E^{(0)}}{E^{(s)}}\right)}
},
\end{equation}
where $E^{(0)}$ denotes the gate error before the QOCT iteration, and $E^{(s)}$ is the saturated gate error after the iteration. Improvement characterizes the order of magnitude of the gate error improved by the QOCT iteration.

\section{Numerical results and discussion}

In principle, we can deal with any environment spectral density
resulting in a bath correlation function that can be expanded 
in the form of a multi-exponential function. Here we consider two
kinds of environment spectral densities or environment correlation
functions: the Lorentzian-like correlation function 
and the Ohmic correlation function.
The Lorentzian spectral density $J_{l}(\omega)=\frac{\alpha}{2\pi}\frac{\gamma^{2}}{(\omega-\Omega)^{2}+\gamma^{2}}$ yields the Lorentzian-like exponential decaying bath correlation function \cite{breuer2002theory,PhysRevA.55.2290,Lorentzian}
\begin{equation}
\label{eq: Lorentz correlation}
{
c_{l}(t-s)=\alpha\frac{\gamma}{2}\exp\left[-\gamma|t-s|-i\Omega(t-s)\right]
}.
\end{equation}
The environment effect is characterized by the correlation strength $\alpha$, the correlation time $\gamma^{-1}$, and the central frequency of the environment spectrum $\Omega$.
The Ohmic correlation function can be derived analytically from the Ohmic spectral density $J_{o}(\omega)=2\alpha_{o}\omega\exp{(-\omega/\omega_{c})}$ as \cite{RevModPhys.59.1}
\begin{equation}
\label{eq: Ohmic correlation}
{
c_{o}(t-s)=2\alpha_{o}\omega_{c}^{2}\left[1+i\omega_{c}(t-s)\right]^{-2}
},
\end{equation} 
where $\alpha_{o}$ is the dimensionless coupling strength and $\omega_{c}$ is the cutoff frequency. Note that function fitting is required to put Eq.~(\ref{eq: Ohmic correlation}) in a multi-exponential form. We present below the numerical results with the parameters in units of $\omega_{0}$ if not stated otherwise.

\subsection{Numerical results}
\begin{figure}[h]
        \subfigure{\includegraphics[width=0.23\textwidth]{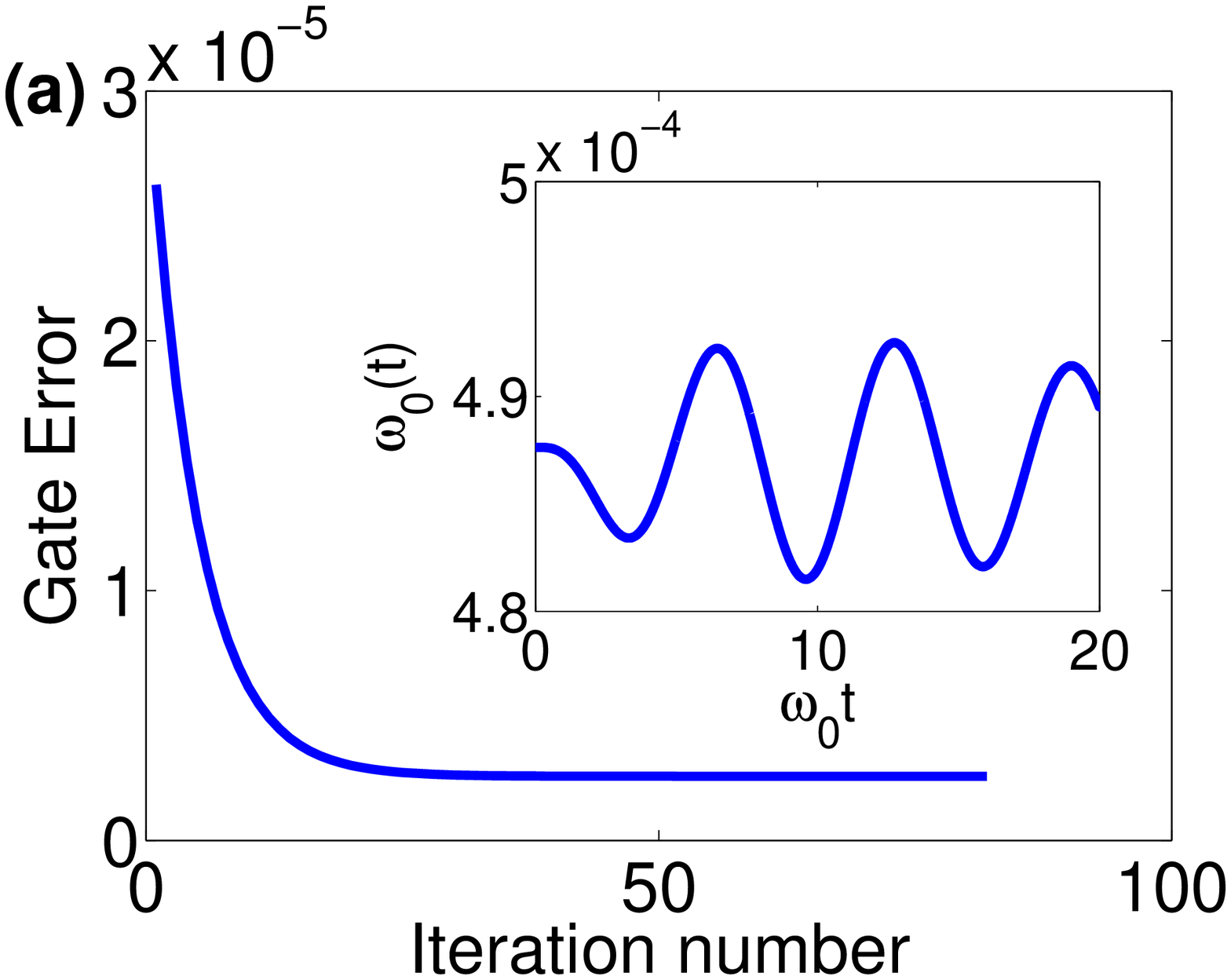}
        }
        \subfigure{\includegraphics[width=0.23\textwidth]{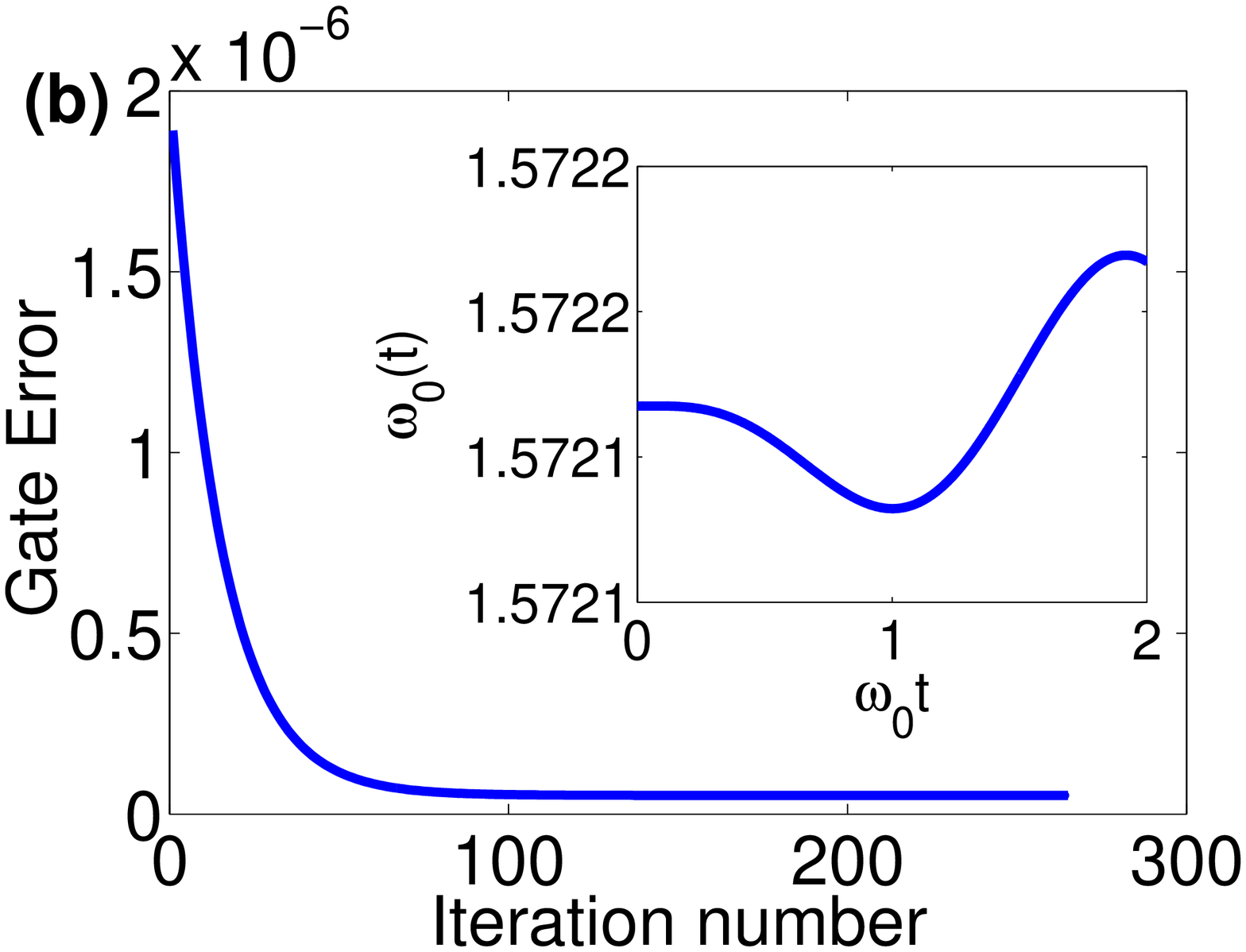}
        }
        \caption{(Color online) Typical QOCT iteration profile and control pulses in
          the Lorentzian-like environment. (a) Identity-gate control
          with $\alpha=0.01$, $\gamma=0.1$, $\Omega=1$, and $t_{f}=20$. (b) $Z$-gate control with $\alpha=0.1$, $\gamma=0.1$, $\Omega=5$, and $t_{f}=2$.}
        \label{fig: iteration and pulse}        
\end{figure}

\begin{figure}[h]
        \subfigure{
                \includegraphics[width=0.23\textwidth]{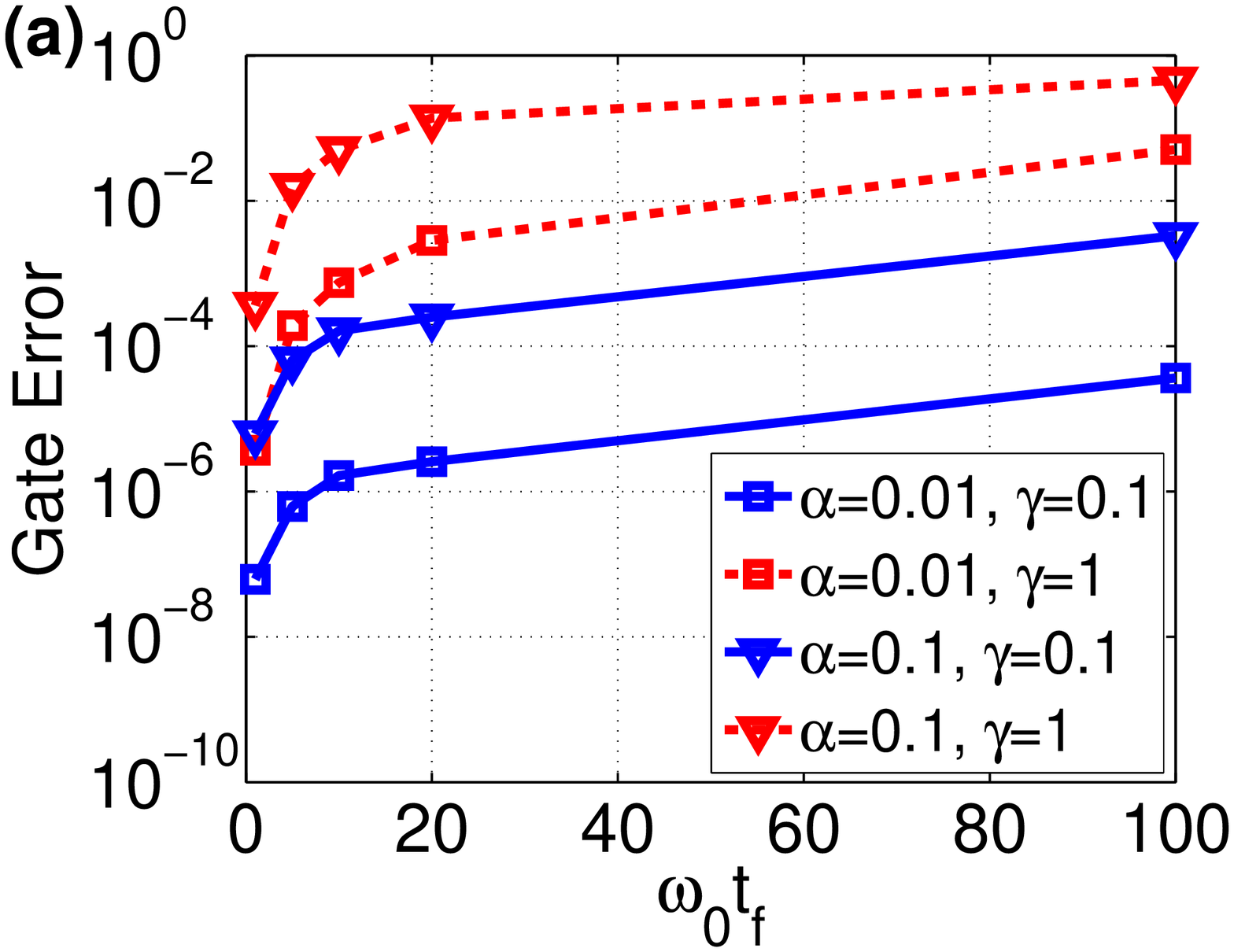}}
        \subfigure{
                \includegraphics[width=0.23\textwidth]{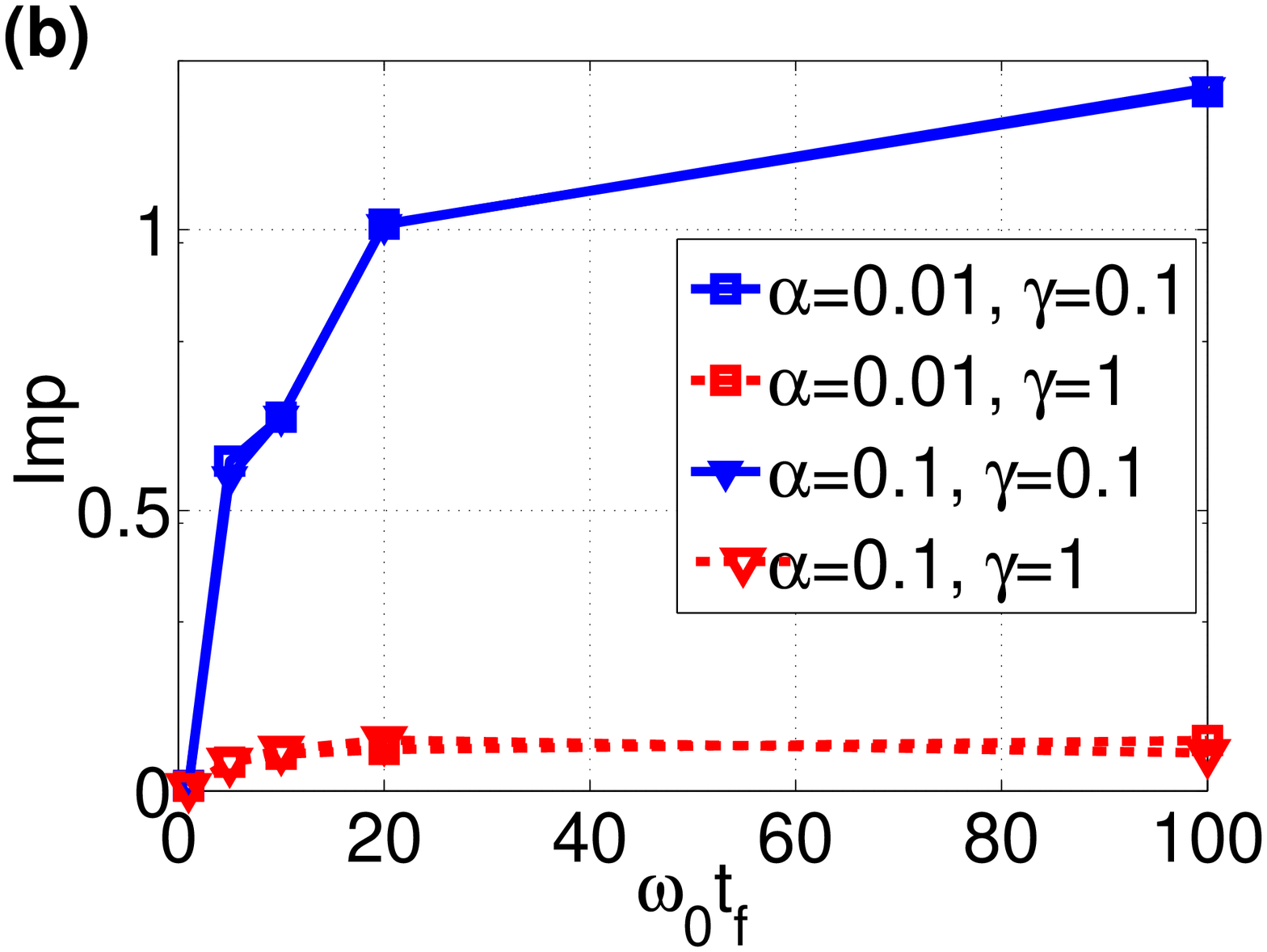}}                
        \subfigure{
                \includegraphics[width=0.23\textwidth]{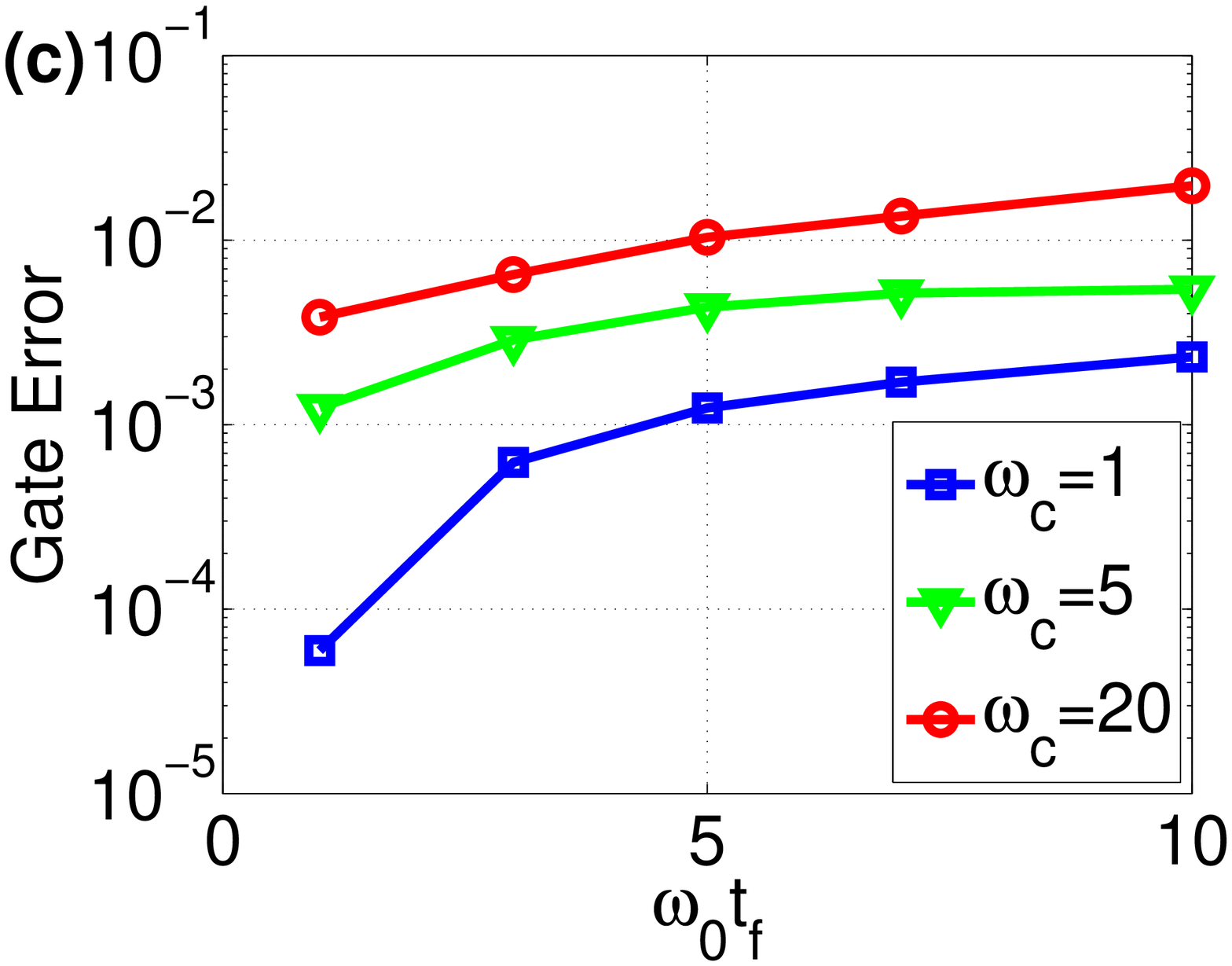}}              
        \subfigure{
                \includegraphics[width=0.225\textwidth]{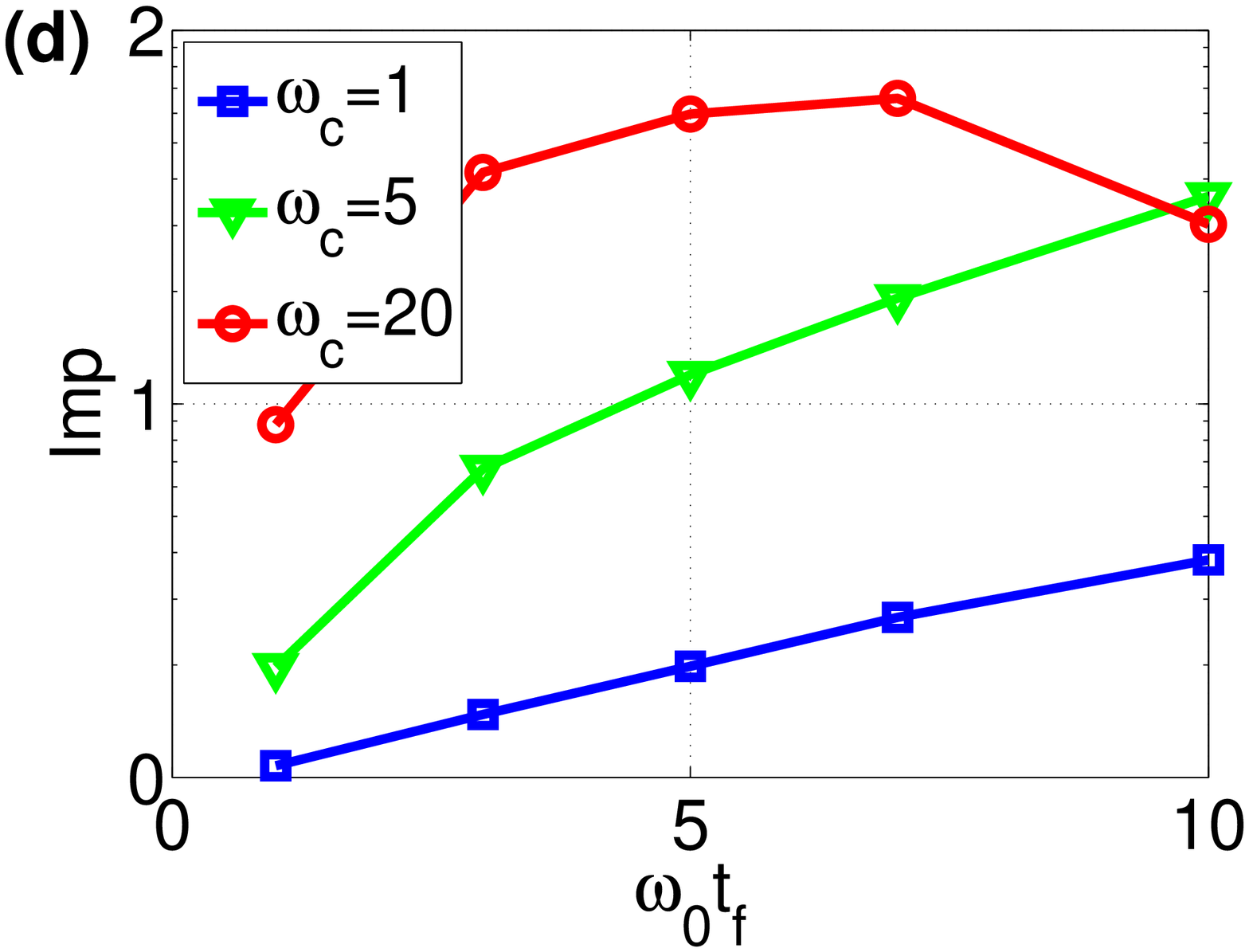}}               
        \caption{\label{fig:I gate} (Color online) Gate error and improvement
          vs gating time of identity-gate control. (a) and (b) correspond
          to a Lorentzian-like environment with $\Omega=1$, and (c)
          and (d)
          correspond to an Ohmic environment with $\alpha_{o}=0.01$.} 
\end{figure}

Figure \ref{fig: iteration and pulse} shows typical optimal pulses (in
the insets) and the monotonic converging behavior of the QOCT
iteration, a favorable feature of the Krotov method, and the
saturation of gate error near the optimal trajectory. The smooth shape
of the optimal control pulses can be easily engineered. Identity gates
serve as quantum memories and thus favor long gating times. Figure
\ref{fig:I gate} shows the gate error after QOCT iteration and
improvement vs gating time of the identity-gate control in both the
Lorentzian-like environment [Figs.~\ref{fig:I gate}(a) and~\ref{fig:I gate}(b)] and the
Ohmic environment [Figs.~\ref{fig:I gate}(c) and~\ref{fig:I gate}(d)]. It appears that
high-fidelity identity gates with error $\mathcal{E}\lesssim10^{-3}$ can be achieved for gating times longer than the system decay time for moderate system decay parameters. Gate control is better performed with weaker qubit-environment coupling strength ($\alpha$ or $\alpha_{o}$ small) and with smaller $\gamma$ or $\omega_{c}$ in both cases. Note that improvement increases as the gating time gets longer. The anomalous crossing in Fig.~\ref{fig:I gate}(d) results from gate error saturation in extreme conditions. A $Z$-gate operation is desired to be fast and thus requires a short gating time. We set a fixed $Z$-gate gating time $t_{f}=2\omega_{0}^{-1}$, which is the smallest multiples of $\omega_{0}^{-1}$ within which an ideal closed system $Z$-gate can be fulfilled in the admissible control range. The results are shown in Tables~\ref{tab: Lorentz Z} and \ref{tab:Ohmic Z}. The trends are similar to the identity-gate control.

The parameters $\gamma$ and $\omega_{c}$ determine the bath
correlation time and the shape of the correlation function. From
Eqs.~(\ref{eq: Lorentz correlation}) and (\ref{eq: Ohmic
  correlation}), it can be shown that larger $\gamma$ or $\omega_c$
corresponds to a bath correlation function of shorter correlation time
and a stronger correlation strength near $s=t$, namely, a relatively
Markovian correlation. In Table~\ref{tab: Lorentz Z}, we demonstrate
the effect of $\Omega$ on $Z$-gate control. The gate error becomes
smaller when $\Omega$ increases. Mathematically, this can be inferred
from Eq.~(\ref{eq: Lorentz correlation}) that large $\Omega$ results
in mutual cancellation of the bath correlation function 
in, for example, the integration of Eq.~(\ref{eq:F}) and thus minor environment effect. Physically, the peak of the spectral density is detuned away from the qubit frequency by large $\Omega$ and results in weak environment-induced decoherence.

\begin{figure*}[t]
        {
        \subfigure{
                \includegraphics[width=0.23\textwidth]{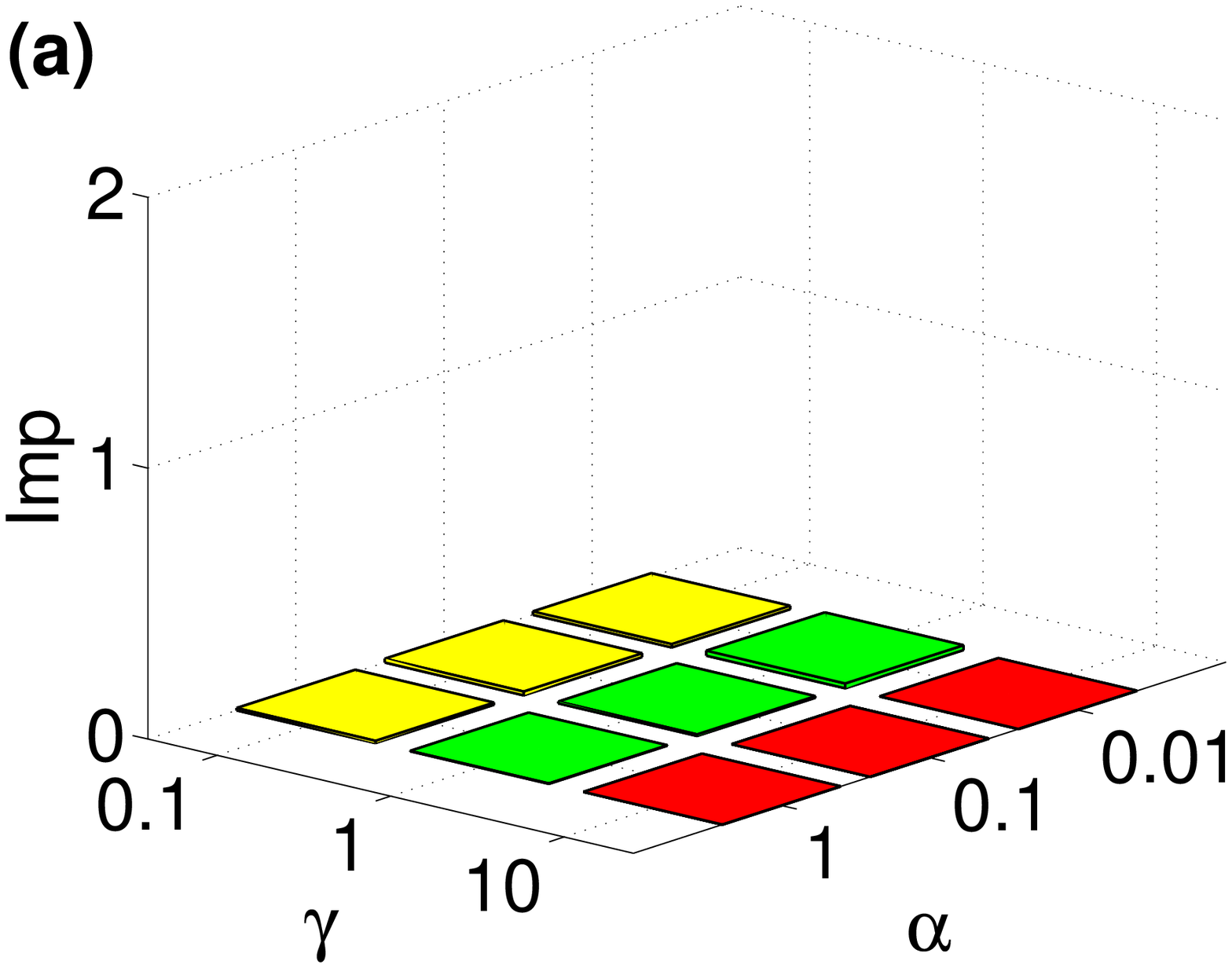}
}
        \subfigure{
                \includegraphics[width=0.23\textwidth]{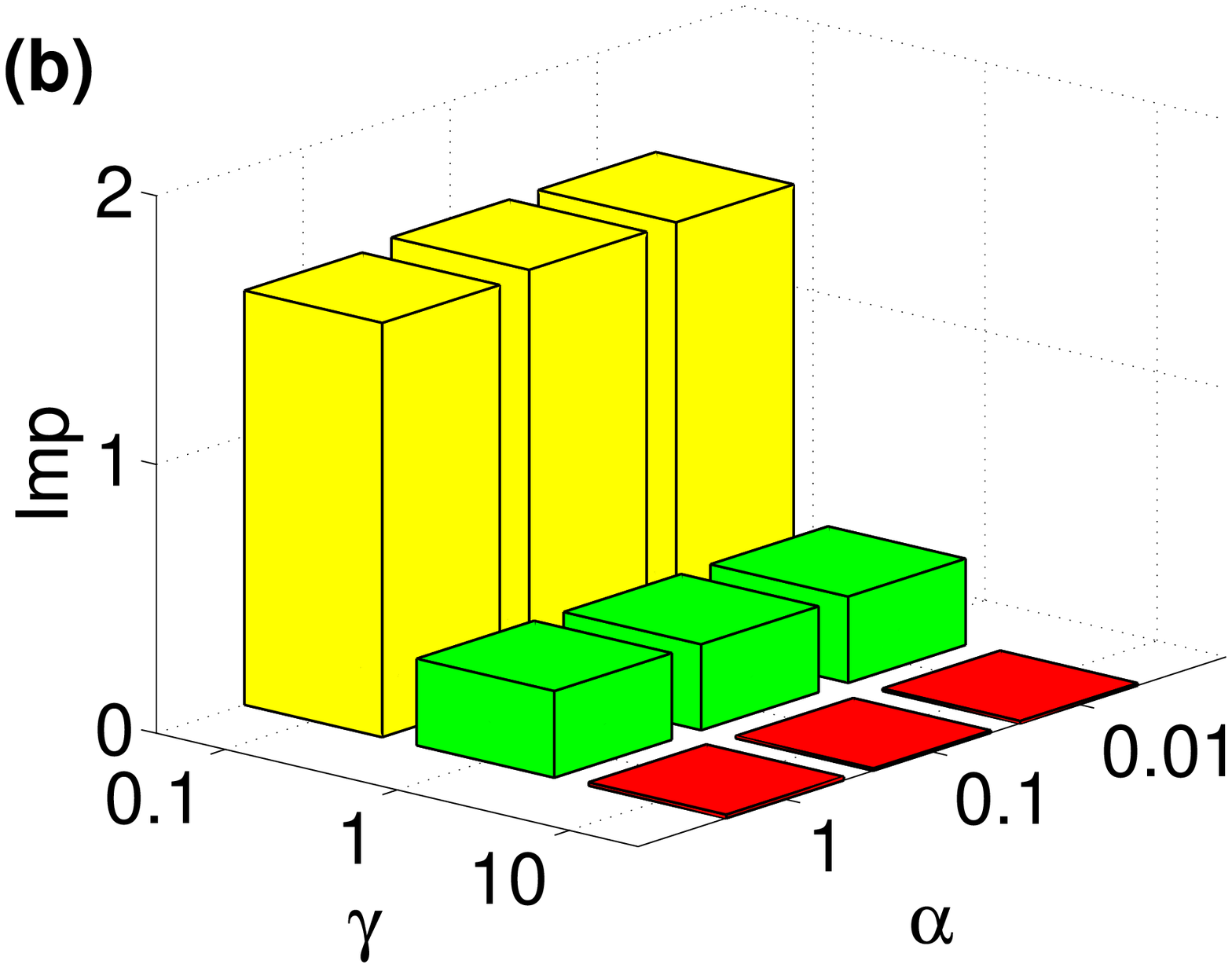}
               }
        \subfigure{
                \includegraphics[width=0.23\textwidth]{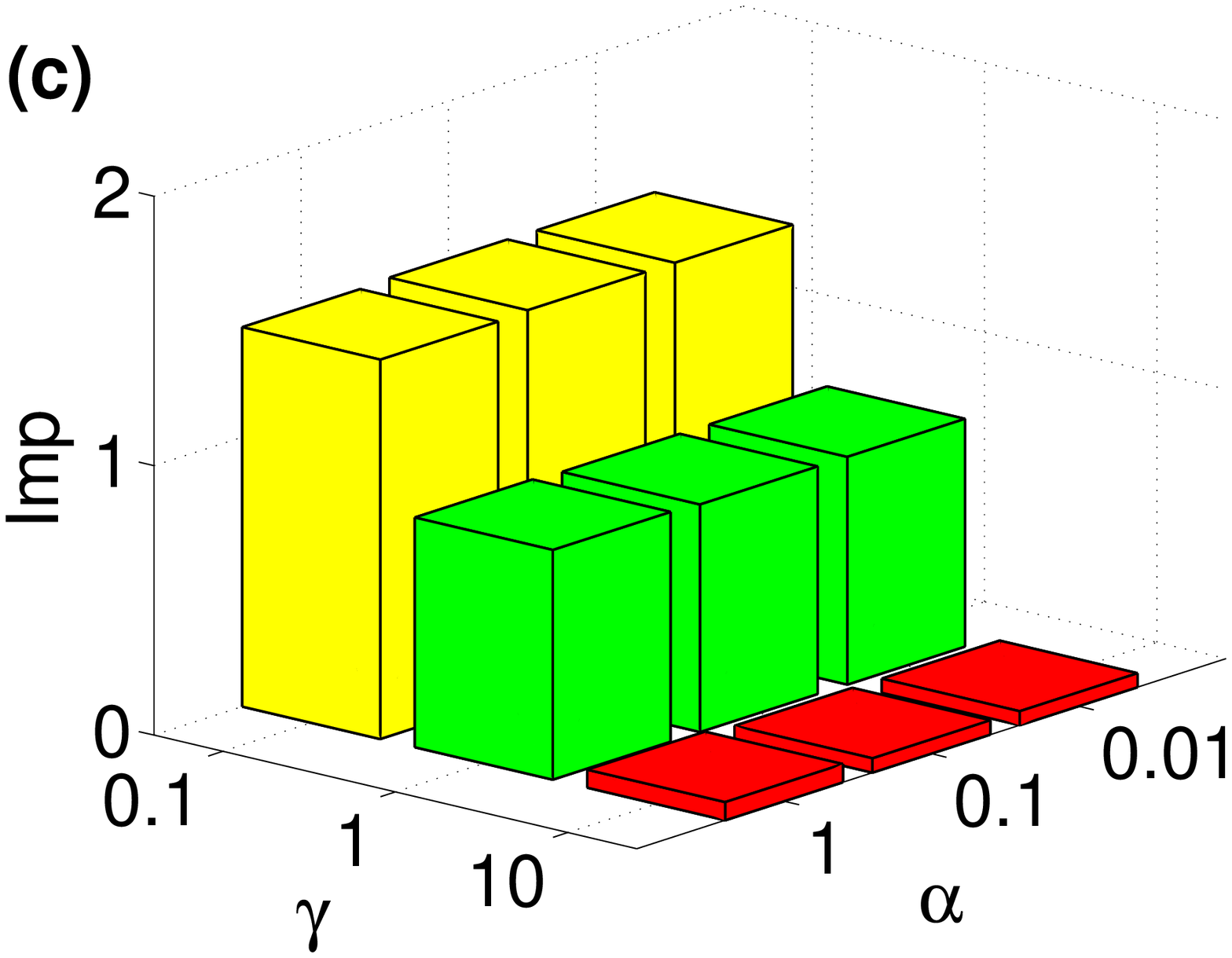}
                }
        \subfigure{
                \includegraphics[width=0.23\textwidth]{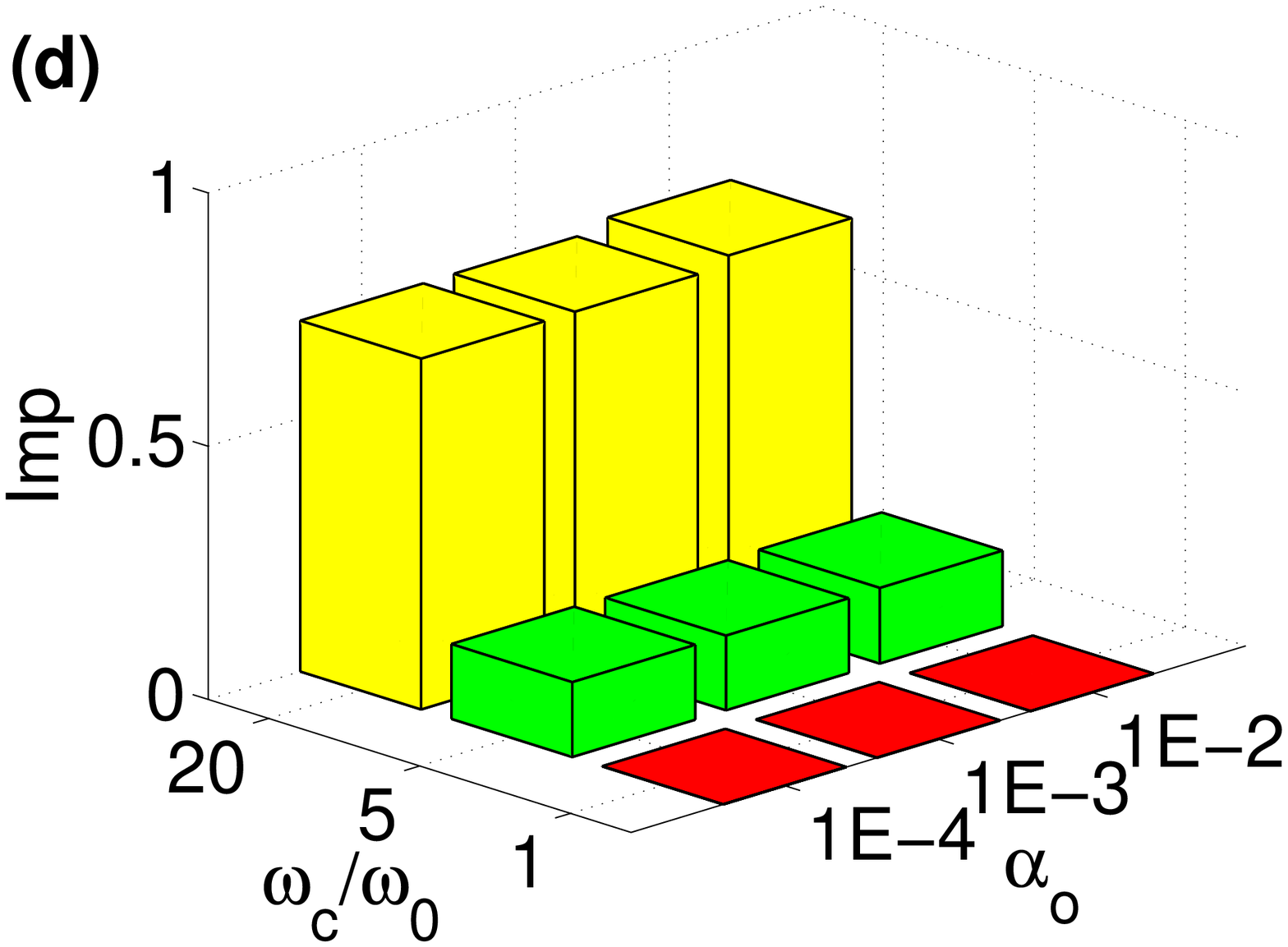}
                }
        \caption{(Color online) Improvement under various conditions of $Z$-gate
          control. (a)--(c) are plotted with
          $\Omega=1$, 5, and 10, respectively, in a Lorentzian-like
          environment and (d) in an Ohmic environment.\label{fig:Z improvement}}
        }
\end{figure*}

Figures~\ref{fig:Z improvement}(a)--\ref{fig:Z improvement}(c) show the plots of the improvement of $Z$-gate
control under various conditions in Lorentzian environment. Apparently
there is hardly any improvement when $\Omega$ is in the vicinity of
the system transition frequency. As the detuning $(\Omega-\omega_{0})$
grows large, we observe great improvement in the $Z$-gate control. The
increase in the improvement is not monotonic. In the Ohmic environment
[Fig.~\ref{fig:Z improvement}(d)], the improvement grows with the cutoff
frequency $\omega_{c}$ as in the identity-gate control
[Fig. \ref{fig:I gate}(d)]. In all cases, neither $\alpha$ nor
$\alpha_{o}$ plays a role in improvement. To study the trend of
improvement, we shall study the exact dynamics and explore the agent
of error correction in the QOCT iteration for the dissipative system we investigate.

\begin{table}[h]
\caption{Errors of the $Z$-gate control in a Lorentzian-like environment under various conditions.}\label{tab: Lorentz Z}
\begin{ruledtabular}
\begin{tabular}{cccc}
\multicolumn{4}{c}{$\Omega=\omega_{0}$}\\
$\alpha$ &$\gamma=0.1$ &$\gamma=1$ &$\gamma=10$\\
\hline
0.01 &$8.89\times10^{-7}$ &$3.53\times10^{-5}$ &$1.10\times10^{-4}$\\
0.1 &$8.81\times10^{-5}$ &$3.31\times10^{-3}$ &$9.57\times10^{-3}$\\
1 &$8.06\times10^{-3}$ &$1.78\times10^{-1}$ &$2.86\times10^{-1}$\\
\hline
\hline
\multicolumn{4}{c}{$\Omega=5\omega_{0}$}\\
$\alpha$ &$\gamma=0.1$ &$\gamma=1$ &$\gamma=10$\\
\hline
0.01 &$5.17\times10^{-10}$ &$1.40\times10^{-6}$ &$9.15\times10^{-5}$\\
0.1 &$5.18\times10^{-8}$ &$1.37\times10^{-4}$ &$7.98\times10^{-3}$\\
1 &$5.35\times10^{-6}$ &$1.10\times10^{-2}$ &$2.59\times10^{-1}$\\
\hline
\hline
\multicolumn{4}{c}{$\Omega=10\omega_{0}$}\\
$\alpha$ &$\gamma=0.1$ &$\gamma=1$ &$\gamma=10$\\
\hline
0.01 &$1.54\times10^{-10}$ &$5.63\times10^{-8}$ &$4.16\times10^{-5}$\\
0.1 &$1.54\times10^{-8}$ &$5.60\times10^{-6}$ &$3.79\times10^{-3}$\\
1 &$1.57\times10^{-6}$ &$5.31\times10^{-4}$ &$1.67\times10^{-1}$\\

\end{tabular}
\end{ruledtabular}
\end{table}

\begin{table}[h!]
\caption{\label{tab:Ohmic Z}
Errors of the $Z$-gate control in an Ohmic environment under various conditions.}
\begin{ruledtabular}
\begin{tabular}{cccc}
$\alpha_{o}$ &$\omega_{c}=1$ &$\omega_{c}=5$ &$\omega_{c}=20$\\
\hline
$10^{-4}$ &$9.73\times10^{-8}$ &$1.97\times10^{-6}$ &$4.64\times10^{-6}$\\
$10^{-3}$ &$9.70\times10^{-6}$ &$1.93\times10^{-4}$ &$4.52\times10^{-4}$\\
$10^{-2}$ &$9.40\times10^{-4}$ &$1.58\times10^{-2}$ &$3.22\times10^{-2}$\\

\end{tabular}
\end{ruledtabular}
\end{table}

\subsection{Conditions for significant improvement}

In the previous section we have shown that in both Lorentzian-like and
Ohmic environments, improvement $\mathcal{I}$ of gate control varies largely
as the qubit-decaying parameters are tuned. We now explore the
conditions under which significant improvement happens and discuss the physics
behind them for the exactly solvable dissipative model we consider.  

The coherence term ($\rho_{21}$, in particular) of the exact solution
to Eq.~(\ref{eq: master equation}) suggests that
$\text{Re}\left[\int_{0}^{t}F(s)ds\right]$ encodes the dissipation
caused by the environment, and
$\text{Im}\left[\int_{0}^{t}F(s)ds\right]$ encodes the phase shift of
the coherence term resulting from the shift in the system frequency due
to the presence of the environment. It reads
\begin{align}
\label{eq: rho21}
\rho_{21}(t)=&\rho_{21}(0)\exp\left(-\int_{0}^{t}\text{Re}\left[F(s)\right]ds\right)\notag\\
&\times\exp\left(i\left(\int_{0}^{t}\omega_{0}(s)ds+\int_{0}^{t}\text{Im}\left[F(s)\right]ds\right)\right)\notag\\
\equiv&\rho_{21}(0)e^{-\kappa}e^{i\phi}.
\end{align} 
The first exponential term represents the dissipation effect
($\kappa$) and the second represents the phase shift ($\phi$). It is
desirable to check how $\kappa$ and $\phi$ behave before and after
the QOCT iteration. A quick check in the exponents of several typical
cases shows that the phase shift
($\phi$) is corrected by the QOCT iteration as shown 
in Table~\ref{tab: phase dissipation}; the
dissipation ($\kappa$), however, can hardly be suppressed. 
This is because in the exactly solvable dissipative model considered
here, the control is only over the $\sigma_z$ term 
that enables the explicit phase correction,
and the control
strength is not very strong ($\left|\epsilon(t)\right|\leq\omega_{0}$) 
for the cases investigated here (note that the dissipation can be
substantially suppressed with large range control  
$\left|\epsilon(t)\right|\leq 20\omega_{0}$ shown in
Sec.~\ref{sec:suppression}).  As a consequence, improvement is
determined by the relative proportion of error that the two effects of the
phase shift and the dissipation contribute to. 
In our dissipative model with control only on the $\sigma_z$ term, if the environment-induced dissipation is the dominant source of gate error, the improvement is limited since after the optimal control iteration, only a minor portion of error can be corrected. In contrast, if the gate error mainly comes from the environment-induced phase shift, then after the optimal control iteration the improvement can be substantial.
\begin{table}[h]
\caption{The phase shift exponent $\phi$ and dissipation exponent
  $\kappa$. The superscripts $(0)$ and $(s)$ indicate the values taken
  before and after the QOCT iteration, respectively.}\label{tab: phase dissipation}
\begin{ruledtabular}
\begin{tabular}{ccccc}

 &Case 1\footnote{Lorentzian-like environment with $\alpha=0.1$,
   $\gamma=0.1$ and $\Omega=5$.} &Case 2\footnote{Same as case 1
   except $\Omega=1$.} &Case 3\footnote{Ohmic environment with
   $\alpha_{o}=0.001$ and $\omega_{c}=20$.} &Case 4\footnote{Same as
   case 3 except $\omega_{c}=1$.}\\
\hline
$\kappa^{(0)}$ &$2.04\times10^{-4}$ &$8.46\times10^{-3}$ &$1.90\times10^{-2}$ &$2.79\times10^{-3}$\\
$\kappa^{(s)}$ &$2.04\times10^{-4}$ &$8.46\times10^{-3}$ &$1.93\times10^{-2}$ &$2.79\times10^{-3}$\\
$\phi^{(0)}$ &$-8.63\times10^{-4}$ &$1.04\times10^{-3}$ &$-2.71\times10^{-2}$ &$-4.61\times10^{-5}$\\
$\phi^{(s)}$ &$-6.66\times10^{-8}$ &$2.91\times10^{-5}$ &$-1.36\times10^{-4}$ &$3.32\times10^{-9}$\\
$\mathcal{I}$ &$1.50$ &$0.0141$ &$0.693$ &$0.000447$
\end{tabular}
\end{ruledtabular}
\end{table}

The dissipation and the phase shift are directly related to the nature of $F(t)$, which is determined by its differential equation, Eq.~(\ref{eq:partialF}). Mathematically, it is possible to find conditions such that the gate error contributed by $\left|\int_{0}^{t_{f}}\text{Im}\left[F(s)\right]ds\right|$ is relatively larger than that by $\left|\int_{0}^{t_{f}}\text{Re}\left[F(s)\right]ds\right|$ and thus determine the conditions for significant improvement. 

Physically, the effect of phase shift and dissipation can be understood as the result of qubit transition frequency shift, namely, Lamb shift, and qubit decay. In the Lorentzian environment with zero detuning, $\Omega=\omega_0$, the spectral density is peaked at and symmetric with respect to $\omega_0$. As the numerical results indicate, the decay rate becomes large and the dissipation effect becomes prominent. On the other hand, the Lamb shift is relatively small due to the symmetric distribution of spectral density with respect to qubit frequency $\omega_0$. In this case the dissipation effect dominates over the phase shift effect, and the improvement due to the optimal control is small. 

However, if the environment central frequency is detuned from the qubit frequency, our numerical results indicate that qubit decay drops dramatically, but the Lamb shift does not change much. This is due to the asymmetric distribution of the spectral density with respect to qubit frequency $\omega_0$ and the result consequently leads to significant improvement. This is in agreement with the trend of improvement observed in the previous sections. Note that this behavior is more prominent as the Lorentzian distribution gets narrower ($\gamma$ small). Similar arguments apply to the Ohmic case. The overall coupling strength $\alpha$ (or $\alpha_{o}$) is irrelevant to the improvement since it does not affect the shape of the spectral density but only the overall value.

\subsection{Suppression of dissipation}
\label{sec:suppression}

So far, we have observed very limited suppression of dissipation
applying QOCT to the two-level dissipative model. This consequence is
model specific, and is due to the control range we specify. In
Eq.~(\ref{eq: rho21}), the control pulse can be designed to directly
cancel the environment-induced phase shift, but can hardly suppress
the dissipation effect through minimizing the magnitude of
$\text{Re}\left[\int_{0}^{t_{f}}[F(s)]ds\right]$. 
However, one can observe in Eq.~(\ref{eq:partialF}) that $\epsilon(t)$ follows the unit imaginary number $i$, so $F(t)$
oscillates faster when the control $\epsilon(t)$ is large in
magnitude. The integral $\int_{0}^{t_{f}}F(s)ds$ is then small in
magnitude due to mutual cancellation. Therefore, if we allow the
optimal control pulse to be considerably large in magnitude compared
to the initial guess and other parameters, the dissipation can be
reduced remarkably after the QOCT iteration. 
A $Z$-gate control with large range control
$\left|\epsilon(t)\right|\leq 20\omega_{0}$ is shown in Fig.~\ref{fig:
  large control}. Note that the gating time is shorter than that of
the small range control,  and furthermore the gate
error is smaller than that of the small ranged control by several 
orders. This is due to both phase shift correction and
dissipation suppression. The question would, however, be whether such
high values of the control strength are physically attainable and
admissible in realistic qubit systems. If so, very high-fidelity gate
operations are practically possible.

\begin{figure}
        \includegraphics[width=0.45\textwidth]{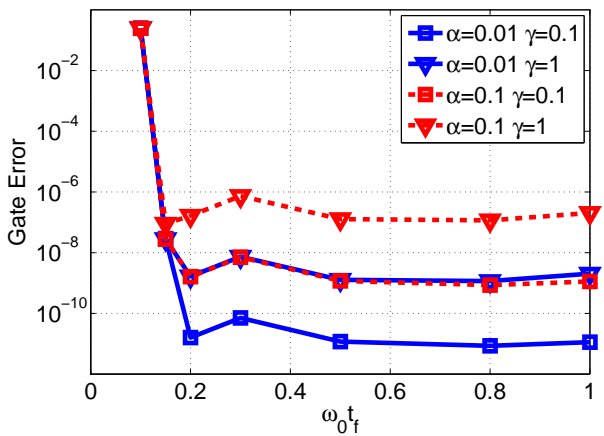}

\caption{(Color online) $Z$-gate control error vs gating time in a
  Lorentzian-like 
  environment with $\Omega=\omega_0$ and large range control $\left|\epsilon(t)\right|\leq20\omega_{0}$. The reduced $\alpha/\omega_{0}(t)$ and $\gamma/\omega_{0}(t)$ ratio is responsible for the much smaller gate error. The evolution time scale is shortened (compared to $\omega_{0}^{-1}$), as well as the gating time.}
        \label{fig: large control}        
\end{figure}


   
\section{Discussion and conclusion}

In this work, we show that an exact open non-Markovian qubit dynamics
can be readily put in the framework of the QOCT to attain single-qubit
gate control.  High-fidelity identity gates and $Z$-gates can be achieved for moderate qubit
decaying parameters with small magnitude control. The optimal pulses
are smooth in shape and easy to implement in experiments.
In cases where the open quantum system models are not
exactly solvable, the perturbative master
equation approaches should be employed for the optimal control solutions
\cite{JPhysB.40.S75,PhysRevLett.102.090401,PhysRevA.78.012358,*PhysRevB.78.165118,*wenin:084504,*PhysRevB.79.224516,*PhysRevA.74.022319,0295-5075-87-4-40003,PhysRevLett.101.010403,PhysRevLett.104.040401,Hwang2012}.
However, 
for models where the exact master
equations are available, our present treatment,  
in contrast to the commonly used perturbation method,
is valid for all orders and free from intrinsic error.

The dissipative model and the QOCT method discussed above can be
readily applied to realistic physical systems such as the circuit QED
system \cite{PhysRevA.69.062320,Wallraff2004,Circuit_QED_review,Sarovar05}
. In circuit QED, the system is realized by a Josephson charge qubit
or a transmon qubit \cite{Koch07} coupled to a coplanar waveguide
resonator and the qubit frequency can be controlled by external
electric voltage and magnetic flux \cite{RevModPhys.73.357,PhysRevA.69.062320,Koch07,Circuit_QED_review}. In principle, this formalism can be applied to any two-level system embedded in a structured environment \cite{PhysRevA.80.012104}, e.g., nitrogen vacancy center in diamond embedded in photonic band-gap \cite{Prawer20062008}.

We introduce the definition of improvement and find that, improvement
is directly related to the mathematical nature of $F(t)$. Physically,
improvement is in close relation to the shape of the spectral density
with respect to the qubit transition frequency. The concept of
improvement does not need to be limited to this specific exactly solvable model, but can also be extended to more general systems that allow no exact solutions. Gaining the insight of improvement, one is able to determine in which condition the improvement is notable, and that applying QOCT to the environment-included open system is necessary. 

In the model (dissipative model) and the control problem $(\sigma_{z}$ control) discussed in our work, the suppression of dissipation is substantial only when we increase the control strength or, in a physically equivalent sense, enlarge the ratio of the qubit frequency to the qubit-environment coupling strength. This result is in agreement with that implicitly stated in \cite{PhysRevA.88.022333} where nonperturbative dynamical decoupling is applied to the same model.

\begin{acknowledgments}
H.S.G. acknowledges support from the
National Science Council in Taiwan under Grant
No. 100-2112-M-002-003-MY3, 
from the National Taiwan University under Grants
No. 103R891400, No. 103R891402 and 102R3253, and
from the
focus group program of the National Center for Theoretical
Sciences, Taiwan.
\end{acknowledgments}

%


\begin{thebibliography}{48}
\bibitem [{Lor()}]{Lorentzian}%
  \BibitemOpen
  \href@noop {} {}\bibinfo {note} {{Note that this transformation is valid only
  when $\Omega$ is sufficiently larger than $\gamma$, the width of the
  Lorentzian. We disregard this condition and, instead, call the function in
  Eq.~(\ref{eq: Lorentz correlation}) the Lorentzian-like correlation
  function.}}\BibitemShut {Stop}%
\end{thebibliography}

\end{document}